\documentclass{article}
\usepackage{icrctc07}

\title{Theory of cosmic ray and $\gamma$-ray production in the supernova
remnant RX~J0852.0-4622 (Vela Jr.)
}
\shorttitle{Cosmic ray production in RX~J0852.0-4622}
\authors{E.G. Berezhko$^{1}$, G. P\"uhlhofer$^{2}$, and H.J. V\"olk$^{3}$.}
\shortauthors{E.G. Berezhko, G. P\"uhlhofer, and H.J. V\"olk}

\afiliations{$^1$Yu.G. Shafer Institute of Cosmophysical Research and Aeronomy,
             31 Lenin Ave., 677980 Yakutsk, Russia \\
             $^2$ Landessternwarte, K\"onigstuhl, 69117 Heidelberg, Germany \\
             $^3$Max-Planck-Institut f\"ur Kernphysik, Postfach 103980, 69029
             Heidelberg, Germany}
\email{berezhko@ikfia.ysn.ru}

\abstract
{Explicitly time-dependent, nonlinear kinetic theory of cosmic ray (CR)
acceleration in supernova remnants (SNRs) has been used to investigate the
properties of the very large SNR RX~J0852.0-4622. The available observations do
not clearly distinguish between a ``nearby'' (at $\sim 200$~pc) and a
``distant'' (at $\sim 1$~kpc) source scenario. Therefore two correspondingly
different models were analyzed. While the 200 pc solution can not be a priory
excluded, the 1 kpc solution turns out to be clearly preferable for physical
reasons. It requires a core collapse supernova (SN) with a massive progenitor
in a molecular cloud $\sim 4000$~yrs ago. The overall synchrotron spectrum and
the filamentary structures in hard X-rays both consistently lead to an
amplified magnetic field $B > 100 \mu$G in the SNR interior. This implies a
suppression of the leptonic TeV $\gamma$-ray emission to about 1 percent of the
flux measured by the H.E.S.S. telescope system which therefore must be
hadronic, consistent with the theoretical solution. Up to the present the 1 kpc
solution has already converted $\sim 10$ percent of the explosion energy into
nonthermal energy, as expected for a Galactic CR source. Also the derived
$\gamma$-ray morphology is consistent with the H.E.S.S. measurements. For the
``nearby'' solution the leptonic and hadronic $\gamma$~ray fluxes are in the
ratio 1:10 which means that this case is also hadronically dominated. However,
the magnetic field strength, consistent with the overall synchrotron spectrum,
differs significantly from that derived from the X-ray filaments. Finally, the
total mechanical energy released amounts to only $1.8 \times 10^{50}$~erg,
uncomfortably low even for a core collapse event.}

\begin{document}
\maketitle


\section{Introduction}

RX~J0852.0-4622 (often, and also here, called Vela Jr.) is a shell-type
supernova remnant (SNR) with a diameter of $2^{\circ}$, located in the Galactic
plane. It was originally discovered in X-rays with ROSAT \cite{Aschenbach}. In
projection Vela Jr. lies entirely within the still much larger Vela SNR and is
only visible in hard X-rays, where the thermal radiation from the Vela SNR is
no longer dominant.  We note that, regarding its size and complexity, Vela
Jr. has similarities with the X-ray SNR RX~J1713.7-3946 also detected in VHE
$\gamma$-ray observations (e.g. \cite{Aha06_1713}). The theoretical analysis of
Vela Jr., summarized below, will also be similar to that for SNR
RX~J1713.7-3946 by \cite{BV06}, and we refer to that paper for more detailed
arguments and references.

The radio emission of Vela Jr. is weak. Only for the northeastern rim a
spectral index can be derived with quite moderate accuracy 
\cite{Duncan-Green}. Vela Jr. was also detected in very high energy (VHE)
$\gamma$-rays by the H.E.S.S. collaboration, at the same flux level as the Crab
Nebula, and its morphology was resolved as a rather circular shell,
e.g. \cite{Aha-Vela07}. Emission from the northwestern rim had been detected
already before by the CANGAROO experiment, e.g. \cite{Enomoto}.

The fairly regular shell-type characteristics of this source have prompted us
to construct a model of the acceleration of both electrons and protons in
detail using an explicitly time-dependent nonlinear kinetic theory of cosmic
ray (CR) acceleration that assumes spherical symmetry \cite{BEK,BV00}. We
emphasize nevertheless that particle injection is not spherically symmetric
which requires a renormalization of the CR energy \cite{VBK03}. The theory
couples particle acceleration on a kinetic level with the gas dynamical
evolution of the system in the aftermath of the SN explosion. However, the
present uncertainties regarding this source are too large as to permit the a
priory-assumption of a unique model. Such important astronomical parameters as
the distance, expansion speed, age, and explosion type are poorly known. It is
not even clear, whether the source is in front or behind the Vela SNR
which itself is generally considered to lie at a distance $d=250 \pm 30$~pc
\cite{Cha}. This led us to consider the construction of two quite different
source scenarios. They correspond to earlier distance estimates: a ``nearby ''
solution with $d=200$~pc \cite{Aschenbach}, and a ``distant'' solution with
$d=1$~kpc \cite{Slane}. To rather different degrees of success these
constructions turn out to be indeed possible. The SNR ages correspond to
$t=1360$~yr and $t=3930$~yr, respectively.

We shall argue that -- for the favored ``distant'' solution -- the observed
nonthermal emission of Vela Jr. indicates that the SNR emerged from a type II
SN explosion into the adiabatic wind bubble of a massive progenitor star. In
this case the major part of the swept-up volume is occupied by the diluted
bubble gas. At the current epoch, however, the SNR shock already propagates
into the increasingly dense shell of ambient interstellar medium (ISM) which
has originally been compressed by the stellar wind. A ``nearby'' scenario, on
the other hand, is only possible for a uniform ambient medium.

\begin{figure}
  \begin{center}
    \includegraphics [width=7.0cm]{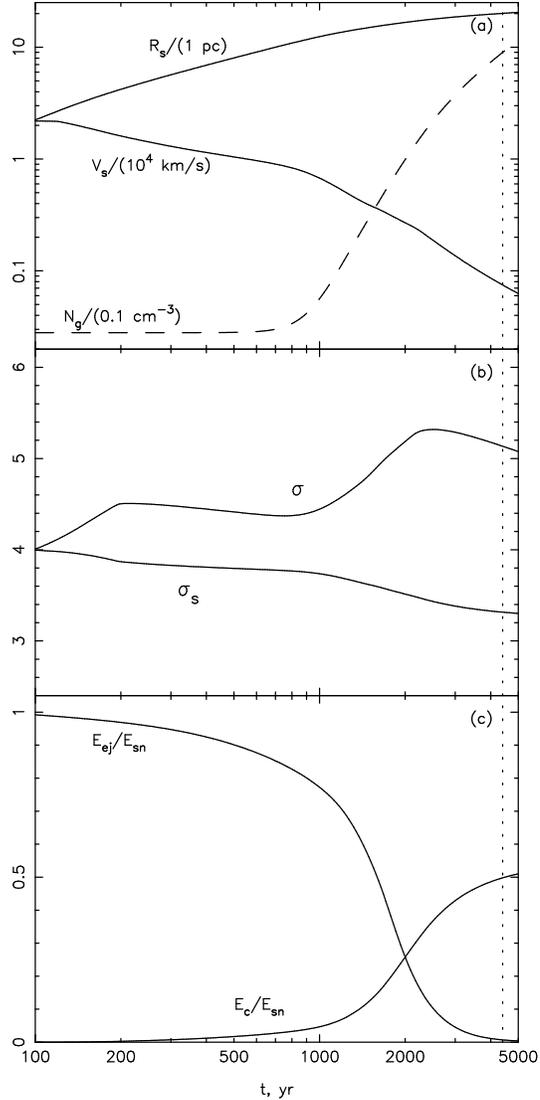}
  \end{center}
  \caption{\label{fig:modelparameters}Calculated hydrodynamic quantities
  for the distanse $d=1$~kpc. 
  (a) Shock radius $R_\mathrm{s}$, velocity $V_\mathrm{s}$, and gas density
      $N_\mathrm{g}$ as functions of time; 
  (b) Total ($\sigma$) and subshock ($\sigma_\mathrm{s}$) compression ratio;
  (c) ejecta energy ($E_\mathrm{ej}$) and CR energy ($E_\mathrm{c}$), where the
  latter has still to be renormalized on account of the lack of spherical
  symmetry. The vertical dotted lines mark the current epoch of SNR evolution.}
\end{figure}

\section{Results}
\subsection{``Distant'' solution}

\begin{figure*}
    \hspace*{0cm}\includegraphics[width=15.0cm]{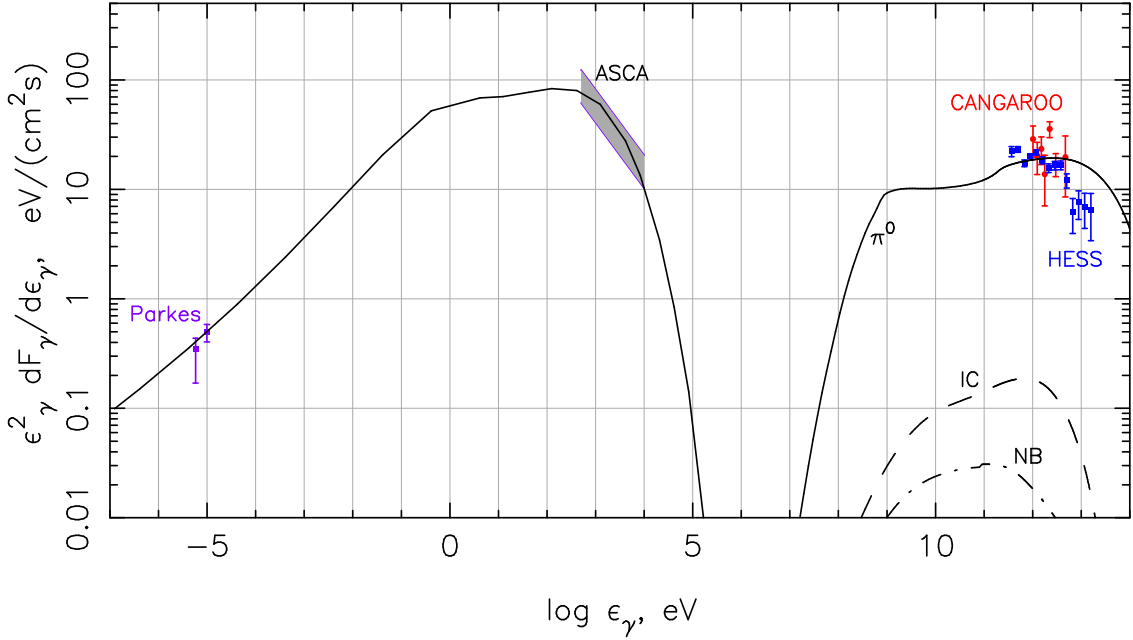}\hfil 
  \caption{\label{fig:spectrum}Calculated spectral energy distribution for Vela
  Jr. for the 1 kpc solution as function of photon energy.
  In the $\gamma$-ray region the solid curve shows the
  $\pi^0$-decay emission, whereas the dashed and the dash-dotted lines denote
  the inverse Compton and nonthermal Bremsstrahlung emissions,
  respectively. Radio \cite{Duncan-Green} and
  X-ray synchrotron fluxes \cite{Slane, Aha-Vela07},
  CANGAROO
  \cite{Enomoto} and H.E.S.S. \cite{Aha-Vela07} TeV data are also
  shown.
}
\end{figure*}

For the ``distant'' solution at 1 kpc the present radius of the SNR blast wave
is $R_\mathrm{s} \approx 17.5$~pc. Such a large size, combined with the need
for a shock that is presently still fast in order to explain the luminosity in
hard X-rays, requires a very low thermal gas density at least in the deeper
interior of the remnant. Therefore the progenitor star must have had originally
a large mass, somewhat below $20 M_{\odot}$, in a surrounding ISM of density
$12 < N_\mathrm{ISM} < 40$~cm$^{-3}$, i.e. in a molecular cloud. Consistent
values for total mechanical energy release and ejected mass are
$E_\mathrm{sn}=2\times 10^{51}$~erg and $M_\mathrm{ej}=3.5 M_{\odot}$,
respectively. The effective magnetic field strength $B$ inside the remnant
should be both consistent with the observation of thin X-ray filaments with the
Chandra telescope \cite{Bamba}, from which we derive a present-day value
$B\approx 130 \mu$G, as well as with the form of the overall, spatially
integrated synchrotron spectrum. To fit the latter we used a value $B=106 \mu$G
-- constant in time -- in satisfactory agreement with the filament value. This
shows that the magnetic field is significantly amplified compared to the value
upstream of the shock, and this is only possible through an effectively
accelerated nuclear CR component. The large B-field at the same time suppresses
the accelerated CR electron component.  To obtain the amplitude of the observed
$\gamma$-ray spectrum a -- theoretically quite plausible -- proton injection
rate $\eta = 3\times 10^{-4}$ is required.

This model and its parameters allow a reasonable fit for the present
hydrodynamical variables like shock radius $R_\mathrm{s}$, shock velocity
$V_\mathrm{s}=750$~km/s, compression ratio and overall CR energy (Fig.1).  We
note that at early times $V_\mathrm{s} \approx 20.000$~km/s is quite large and
the (central) gas density ($N_\mathrm{g} \approx 0.003$~cm$^{-3}$~very low,
whereas in the swept-up shell of molecular cloud gas the shock has finally
strongly decelerated, being already far beyond sweep-up. Not taking escape of
the highest energy particles into account over this recent phase, the maximum
proton energies are $p_{\mathrm {max}}\approx 7\times 10^5$~GeV. Had we taken
an effective B-field strength $\propto (N_{\mathrm g} V_s)^{1/2}$, with the
above value characterizing the present epoch, then the maximum momentum would
be even higher. The spectral energy density (Fig. 2) for the 1 kpc model is
characterized by a ``flat-top'' synchrotron peak due to synchrotron cooling,
together with a hadronic dominance of the $\gamma$-ray emission spectrum by
roughly two orders of magnitude. The synchrotron losses as a result of the
amplified B-field permit a good fit to the X-ray data. The gamma-ray data can
be understood in terms of particle escape, despite the fact that the magnetic
field value was taken constant during SNR evolution. Otherwise the discrepancy
in the cutoff energy would be even somewhat greater. Disregarding this
difficulty for the moment, the hadronic dominance is a robust result, based not
only on the synchrotron spectrum but on the X-ray morphology as well. The
(renormalized) energy in nonthermal particles at the present epoch amounts to
$\approx 10$~percent of the total mechanical energy $E_\mathrm{sn}= 2\times
10^{51}$~erg released in the SN explosion. Therefore from the point of view of
energetics this solution for Vela Jr. fulfills the average requirement on a SNR
source of the Galactic CRs.

\begin{figure}
  \begin{center} 
   \includegraphics [width=7.0cm]{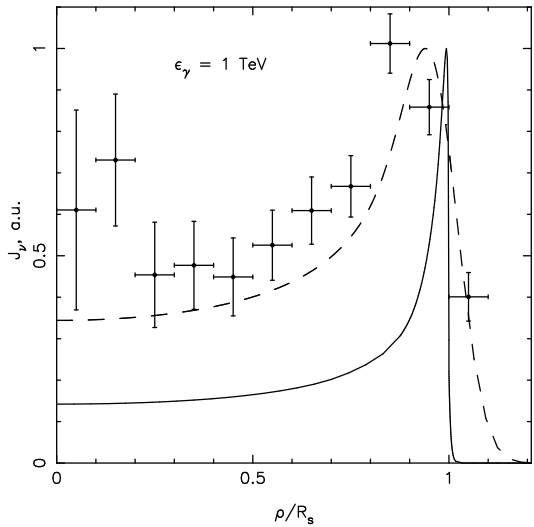} 
   \end{center}
\caption{\label{fig:morphology} The $\gamma$-ray emissivity for the
  energy $\epsilon_{\gamma}=1$~TeV as function of projected, normalized
  radial distance $\rho/R_\mathrm{s}$ for the ``distant'' solution. The
  calculated radial profile is represented by the solid line. Data points are
  from the Northern part of Vela Jr. \cite{Aha-Vela07}, with an analysis point
  spread function of Gaussian width $0.06^{\circ}$. The dashed line represents
  the calculated profile convolved with the same point spread function.}
\end{figure}

The $\gamma$-ray shell morphology at TeV energies with an observed
center-to-limb intensity ratio of $\sim 0.35$, also agrees reasonably well with
the model (Fig.3), given the limited angular resolution of the instrument, and
ignoring the two data points in the central region as possibly due to a central
SNR component. It is worthwhile to comment that the inferred spherically
symmetric 3-dim. thickness of the $\gamma$-ray shell is much smaller and
corresponds to only about 1 percent of the shock radius!

\subsection{``Nearby'' solution}
 
The ``nearby'' solution represents a much earlier stage of SNR
evolution. Although it cannot be excluded right away, the spectrum and the
morphology in the TeV $\gamma$-ray region can only be fitted with more liberal
criteria. Also the magnetic field strengths, derived from the filamentary X-ray
morphology on the one hand, and on the overall synchrotron spectrum on the
other, differ significantly. Finally, the total mechanical energy release
$E_\mathrm{sn}\approx 1.8 \times 10^{50}$~erg is uncomfortably low, even though
this may not be impossible \cite{Pastorello}. In conclusion, everything argues
for the ``distant'' solution. A strict empirical proof in favor of this
solution would come from observations of the true distance of the SNR. The
hadronic dominance of the $\gamma$-ray emission is, however, independent of
either one of these locations of the source.

This work has been supported by the Russian Foundation for Basic Research
(grants 05-02-16412, 06-02-96008, 07-02-0221).


\bibliography{icrc0597rev3}

\begin{thebibliography}{10}

\bibitem{Asch98}
B. {Aschenbach}.
\newblock {\em Nature}, 396:141, 1998.

\bibitem{Aha06}
F.A. {Aharonian et al.(HESS Collaboration)}.
\newblock {\em Astron. Astrophys.}, 449:223, 2006.

\bibitem{BV06}
E.G. {Berezhko} and H.J. {V\"olk}.
\newblock {\em Astron. Astrophys.}, 451:981, 2006.

\bibitem{DG00}
A.R. {Duncan} and D.A. {Green}.
\newblock {\em Astron. Astrophys.}, 364:732, 2000.

\bibitem{Aha07}
F.A. {Aharonian et al. (HESS Collaboration)}.
\newblock {\em Astrophys. J.}, 661:236, 2007.

\bibitem{Eno06}
R. {Enomoto et al.}.
\newblock {\em Astrophys. J.}, 652:1268, 2006

\bibitem{Ber96}
E.G. {Berezhko et al.}.
\newblock {\em JETPh}, 82:1, 1996.

\bibitem{BV00}
E.G. {Berezhko} and H.J. {V\"olk}.
\newblock {\em Astron. Astrophys.}, 375:183, 2000.

\bibitem{vbk05}
H.J. {V\"olk et al.}.
\newblock {\em Astron. Astrophys.}, 409:536, 2005.

\bibitem{Cha99}
A.N. {Cha et al.}.
\newblock {\em Astrophys. J.}, 515:L25, 1999.

\bibitem{Sla01}
P. {Slane et al.}.
\newblock {\em Astrophys J.}, 548:814, 2001.

\bibitem{Bamab}
A. {Bamba et al.}.
\newblock {\em Astrophys. J.}, 632:294, 2005.

\bibitem{Pastorello}
A. {Pastorello et al.}.
\newblock {\em MNRAS}, 347:74, 2004.

\end{thebibliography}
\bibliographystyle{unsrt}

\end{document}